\documentclass[12pt]{iopart}

\usepackage{iopams}
\usepackage[english]{babel}
\usepackage{graphicx}
\usepackage{cite}
\usepackage{color}
\usepackage{bbm}

\begin{document}

\title[]{Cross-cavity quantum Rabi model}

\author{C. Huerta Alderete}
\address{Instituto Nacional de Astrof\'{\i}sica, \'Optica y Electr\'onica, Calle Luis Enrique Erro No. 1, Sta. Ma. Tonantzintla, Pue. CP 72840, M\'exico}

\author{B. M. Rodr\'iguez-Lara}
\address{Instituto Nacional de Astrof\'{\i}sica, \'Optica y Electr\'onica, Calle Luis Enrique Erro No. 1, Sta. Ma. Tonantzintla, Pue. CP 72840, M\'exico}
\ead{bmlara@inaoep.mx}

\begin{abstract}
We introduce the cross-cavity quantum Rabi model describing the interaction of a single two-level system with two orthogonal boson fields and propose its quantum simulation by two-dimensional, bichromatic, first-sideband driving of a single trapped ion.
We provide an introductory survey of the model, including its diagonalization in the two-level system basis, numerical spectra and its characteristics in the weak, ultra strong and deep strong coupling regimes.
We also show that the particular case of degenerate field frequencies and balanced couplings allows us to cast the model as two parity deformed oscillators in any given coupling regime.
\end{abstract}

\pacs{42.50.-p,03.67.Lx}

\maketitle

\section{Introduction}

The Rabi model \cite{Rabi1936p324,Rabi1937p652} is an integrable model describing the interaction of atomic angular momentum with an external classical magnetic field.
The introduction of a quantum field instead of a classical field  produces the quantum Rabi model (QRM); e.g.
the interaction of just the single neutral atom with a quantum field, under minimal coupling, the long wave and the two-level approximations, instead of a collection of them \cite{Dicke1954p99}.
In the weak coupling regime, which happens in standard experiments where the coupling parameter is small compared to the field frequency, a rotating wave approximation (RWA) can be implemented and the QRM becomes the Jaynes-Cummings model (JCM). This was the first version of the QRM to be analytically solved \cite{Jaynes1963p89}.
The validity of the RWA is broken as the coupling strength grows and the full QRM remained unsolvable for any given coupling strength to field frequency ratio until recently \cite{Braak2011p100401}.
The original and complementary proposals for an analytic solution \cite{Chen2012p023822,Braak2013p23,Braak2013p175301} and a series of proposals to realize the QRM in different experimental  platforms, both quantum \cite{Todorov2009p186402,Niemczyk2010p772,Ballester2012p021007,KenaCohen2013p827,Askenazi2014p043029,Scalari2012p1323,Muravev2011p075309,Pedernales2015p15472} and classical \cite{Crespi2012p163601,RodriguezLara2013p12888,RodriguezLara2014p1784,RodriguezLara2014p1719}, have rekindled the interest on the QRM \cite{Wolf2012p053817,Zhong2013p415302,Gardas2013p265302,Moroz2013p319,Deng2013p224018,Moroz2014p252,Maciejewski2014p16}.

\begin{figure}
\centering \includegraphics[scale= 1]{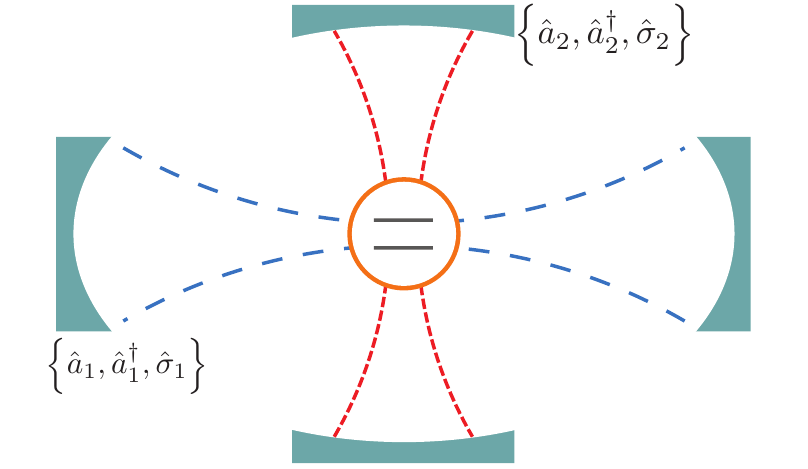}
\caption{(Color online) Sketch of the cross-cavity Rabi model, two orthogonal field modes interact under minimal coupling and long wavelength approximation with one two-level system.} \label{fig:Fig1}
\end{figure}

While extensions for the QRM where the number of qubits  \cite{Braak2013p224007,Peng2012p365302,Chilingaryan2013p335301,Peng2014p265303,Duan2015p121,Wang2014p54001} or fields \cite{Zhang2013p102104,Chilingaryan2015p245501,Duan2015p34003} are increased have been studied in the literature,
here, we want to focus in one configuration that might prove interesting.
Let us imagine a two-level atom interacting with the fields of two cavities in an orthogonal configuration, under minimal coupling and the long wavelength approximation, we can arrive to what we will call a cross-cavity quantum Rabi model,
\begin{eqnarray} \label{eq:CCQRM}
\hat{H} = \frac{1}{2} \omega_{0} \hat{\sigma}_{0} + \sum_{j=1}^{2} \omega_{j} \hat{a}_{j}^{\dagger} \hat{a}_{j} + \sum_{j=1}^{2}  g_{j} \left( \hat{a}_{j}^{\dagger} + \hat{a}_{j} \right) \hat{\sigma}_{j},
\end{eqnarray}
where the qubit has an energy gap $\omega_{0}$ and is described by the Pauli matrices, $\hat{\sigma}_{j}$ with $j=0,1,2$, the boson fields have frequencies $\omega_{j}$ and are described by the annihilation (creation) operators, $\hat{a}_{j}$ ($\hat{a}^{\dagger}_{j}$) with $j=1,2$, and the strength of their couplings is provided by the parameters $g_{j}$ with $j=1,2$.
Figure \ref{fig:Fig1} shows an sketch of our gedankenexperiment where each of the orthogonal fields interact with its corresponding dipole component of the two-level system.
This effective model can be related to the vibrational modes of a single polyatomic molecule interacting with an external magnetic field under linear Jahn-Teller coupling in just two dimensions instead of three \cite{LonguetHiggins1957p425,RomeroRochin1989p6103} and to the Raman adiabatic driving of a single four-level atom coupled to two cavity electromagnetic field modes \cite{Fan2014p023812}.
Note that time evolution in the restricted case of weak couplings and fields of equal frequencies has been given in the literature \cite{GarciaMelgarejo2013,GarciaMelgarejo2013p31}.

In the following, we will propose a trapped-ion quantum simulation of our cross-cavity QRM.
We will diagonalize the Hamiltonian in the qubit basis following a Fulton-Gouterman approach and show that the case of degenerate fields and balanced couplings reduces to the study of two parity deformed oscillators for any given coupling parameter.
For the sake of completeness, we will also provide numerical spectra for the general model in the weak and deep-strong coupling regimes.

\section{Trapped-ion quantum simulation}

In order to leave the gedankenexperiment behind and experimentally motivate the study of our cross-cavity QRM, let us extend the recent proposal to simulate the standard QRM with trapped ions \cite{Pedernales2015p15472}.
We will use two orthogonal pairs of bichromatic driving fields instead of just one,
\begin{eqnarray}
\hspace{-1.75cm} \hat{H}_{TI} = \frac{1}{2} \omega_{0} \hat{\sigma}_{0} + \sum_{j=1}^{2} \left\{  \nu_{j} \hat{a}^{\dagger}_{j} \hat{a}_{j} + \sum_{k=-1,+1} \Omega_{j,k} \cos \left[ \eta_{j,k} \left( \hat{a}_{j}^{\dagger} + \hat{a}_{j} \right) - \omega_{j,k} t + \phi_{j,k} \right] \hat{\sigma}_{j} \right\}.
\end{eqnarray}
Here, the two-level trapped-ion is described by the energy gap $\omega_{0}$ and the Pauli matrices $\hat{\sigma}_{j}$ with $j=0,1,2$, the quantized center of mass vibration modes by the mechanical oscillation frequencies $\nu_{j}$ and the annihilation (creation) operators $\hat{a}_{j}$ ($\hat{a}_{j}^{\dagger}$) with $j=1,2$, and we have two red- and blue-detuned, $k = -1$ and $k=1$ in that order, classical driving lasers with frequencies tuned to the first sideband transitions plus some small detuning $\delta_{j,k}$,
\begin{eqnarray}
\omega_{j,k} = \omega_{0} + k \nu_{j} + \delta_{j,k},
\end{eqnarray}
these classical fields have associated Lamb-Dicke parameters $\eta_{j,k}$, phases $\phi_{j,k}$, and couple to their corresponding dipole components with strength $\Omega_{j, k}$.
Expressing the trigonometric functions in exponential form, using the disentangling property $e^{ (\alpha \hat{a}^{\dagger} + \alpha^{\ast} \hat{a})} = e^{- \frac{1}{2}\vert \alpha \vert^2} e^{\alpha \hat{a}^{\dagger}} e^{\alpha^{\ast} \hat{a}}$, the power series expansion of the exponential, and moving into the rotating frame defined by the uncoupled part of the Hamiltonian,
\begin{eqnarray}
\hat{H}_{0} =\frac{1}{2} \omega_{0} \hat{\sigma}_{0} + \sum_{j=1}^{2}  \nu_{j} \hat{a}^{\dagger}_{j} \hat{a}_{j},
\end{eqnarray}
yields an effective interaction Hamiltonian after some manipulation,
\begin{eqnarray}
\hat{H}_{I} &=& \sum_{j,k} \frac{(-i)^{j-1}}{2} \Omega_{j,k} e^{- \frac{1}{2} \vert \eta_{j,k} \vert^{2}} \times \nonumber \\
&&  \left[ \sum_{p,q=0}^{\infty} \frac{(- i \eta_{j,k})^{p}(- i \eta_{j,k})^{q}}{p!q!} \hat{a}^{\dagger p}_{j} \hat{a}_{j}^{q} e^{i \left[  2 \omega_{0} + \left( q-p+k \right) \nu_{j} + \delta_{j,k}  \right] t} e^{-i \phi_{j,k} } + \right. \nonumber \\
&& \left. \sum_{r,s=0}^{\infty} \frac{(i \eta_{j,k})^{r}(i \eta_{j,k})^{s}}{r!s!} \hat{a}^{\dagger r}_{j} \hat{a}_{j}^{s} e^{i \left[ \left( r-s-k \right) \nu_{j} - \delta_{j,k}  \right] t} e^{i \phi_{j,k}}    \right] \hat{\sigma}_{+}    +  \mathrm{h.c}.
\end{eqnarray}
Typically, the ion energy gap is in the optical region and large compared to all other parameters, $\omega_{0} \gg \nu_{j}, \delta_{j,k}, \Omega_{j,k}$, thus the terms oscillating at optical frequencies, $ 2 \omega_{0} + \left( q-p+k \right) \nu_{j} + \delta_{j,k}  - \phi_{j,k}$, will average to zero in any realistic measurement scenario.
This allows us to focus on the approximate effective interaction Hamiltonian,
\begin{eqnarray}
\hspace{-1.75cm}\hat{H}_{I} &\approx& \sum_{j,k,r,s} \frac{(-i)^{j-1}}{2} \Omega_{j,k} e^{-\frac{1}{2} \vert \eta_{j,k} \vert^{2}} \frac{(i \eta_{j,k})^{r}(i \eta_{j,k})^{s}}{r!s!} \hat{a}^{\dagger r}_{j} \hat{a}_{j}^{s} e^{i \left[ \left( r-s-k \right) \nu_{j} - \delta_{j,k}  \right] t} e^{i \phi_{j,k}} \hat{\sigma}_{+}    +   \mathrm{h.c.}
\end{eqnarray}
After this optical RWA, we can do a mechanical RWA where those terms rotating at frequencies proportional to $\nu_{j}$ average to zero if and only if the mechanical vibration frequency is larger than the first sideband detunings and coupling strengths, $\nu_{j} \gg \delta_{j,k}, e^{-\frac{1}{2}\vert \eta_{j,k} \vert^{2}} \Omega_{j,k}$.
After some manipulation, we can write the following,
\begin{eqnarray}
\hspace{-1.75cm} \hat{H}_{I} = i \frac{\eta_{1,-1}}{2} \Omega_{1,-1} e^{- \frac{1}{2} \vert \eta_{1,-1} \vert^{2}} \left[ \frac{\hat{a}_{1}^{\dagger} \hat{a}_{1}!}{(\hat{a}_{1}^{\dagger} \hat{a}_{1} + 1)!}  L_{\hat{a}_{1}^{\dagger}\hat{a}_{1}}^{(1)}(\vert \eta_{1,-1} \vert^{2}) \hat{a}_{1} \hat{\sigma}_{+} e^{-i \delta_{1,-1} t} e^{i \phi_{1,-1}} + h.c.   \right] + \nonumber \\
\hspace{-.85cm} + i \frac{\eta_{1,1}}{2} \Omega_{1,1} e^{- \frac{1}{2} \vert \eta_{1,1} \vert^{2}} \left[ \hat{a}_{1}^{\dagger}  \frac{\hat{a}_{1}^{\dagger} \hat{a}_{1}!}{(\hat{a}_{1}^{\dagger} \hat{a}_{1} + 1)!}   L_{\hat{a}_{1}^{\dagger}\hat{a}_{1}}^{(1)}(\vert \eta_{1,1} \vert^{2})  \hat{\sigma}_{+} e^{-i \delta_{1,1} t} e^{i \phi_{1,1}} + h.c.   \right] + \nonumber \\
\hspace{-.85cm} + \frac{\eta_{2,-1}}{2} \Omega_{2,-1} e^{- \frac{1}{2} \vert \eta_{2,-1} \vert^{2}} \left[ \frac{\hat{a}_{2}^{\dagger} \hat{a}_{2}!}{(\hat{a}_{2}^{\dagger} \hat{a}_{2} + 1)!}  L_{\hat{a}_{2}^{\dagger}\hat{a}_{2}}^{(1)}(\vert \eta_{2,-1} \vert^{2}) \hat{a}_{2} \hat{\sigma}_{+} e^{-i \delta_{2,-1} t} e^{i \phi_{2,-1}} + h.c.   \right] + \nonumber \\
\hspace{-.85cm} + \frac{\eta_{2,1}}{2} \Omega_{2,1} e^{- \frac{1}{2} \vert \eta_{2,1} \vert^{2}} \left[ \hat{a}_{2}^{\dagger}  \frac{\hat{a}_{2}^{\dagger} \hat{a}_{2}!}{(\hat{a}_{2}^{\dagger} \hat{a}_{2} + 1)!}   L_{\hat{a}_{2}^{\dagger}\hat{a}_{2}}^{(1)}(\vert \eta_{2,1} \vert^{2})  \hat{\sigma}_{+} e^{-i \delta_{2,1} t} e^{i \phi_{2,1}} + h.c.   \right], \nonumber \\
\end{eqnarray}
where the function $L_{n}^{(m)}(x)$ is a generalized Laguerre polynomial.
At this point, we should note that working in the Lamb-Dicke regime, $\eta_{j,k} \sqrt{\langle \hat{a}_{j}^{\dagger} \hat{a}_{j} \rangle} \ll 1 $, and an adequate choice of parameters,
\begin{eqnarray}
  \phi_{1,-1} = \phi_{1,1} =\phi_{2,1} = - \frac{\pi}{2}, \qquad \phi_{2,-1} = \frac{\pi}{2}, \qquad \nonumber \\
  \eta_{j,\pm1} \Omega_{j, \pm1} e^{- \frac{1}{2}\vert \eta_{j,\pm 1} \vert^{2}} = \eta_{j,\mp1} \Omega_{j, \mp1} e^{- \frac{1}{2}\vert \eta_{j,\mp 1} \vert^{2}},
\end{eqnarray}
allows us to write,
\begin{eqnarray}
\hspace{-1.75cm} \hat{H}_{I} \approx  g_{1} \left[ \hat{a}_{1}^{\dagger} e^{\frac{i}{2} \left( \delta_{1,-1} - \delta_{1,1} \right) t} +  \hat{a}_{1} e^{- \frac{i}{2} \left( \delta_{1,-1} - \delta_{1,1} \right) t} \right] \left[ \hat{\sigma}_{+} e^{-\frac{i}{2} \left( \delta_{1,-1} + \delta_{1,1} \right) t} +  \hat{\sigma}_{-} e^{ \frac{i}{2} \left( \delta_{1,-1} - \delta_{1,1} \right) t}  \right] + \nonumber \\
\hspace{-.85cm} + g_{2} \left[ \hat{a}_{2}^{\dagger} e^{\frac{i}{2} \left( \delta_{2,-1} - \delta_{2,1} \right) t} +  \hat{a}_{2} e^{- \frac{i}{2} \left( \delta_{2,-1} - \delta_{2,1} \right) t} \right] \left[ \hat{\sigma}_{+} e^{-\frac{i}{2} \left( \delta_{2,-1} + \delta_{2,1} \right) t} +  \hat{\sigma}_{-} e^{ \frac{i}{2} \left( \delta_{2,-1} + \delta_{2,1} \right) t}  \right]. \nonumber \\
\end{eqnarray}
We can make another transformation and move into the uncoupled rotating frame,
\begin{eqnarray}
\hat{H}^{\prime} = - \frac{1}{4} \left( \delta_{1,-1} + \delta_{1,1} \right) \hat{\sigma}_{0} + \frac{1}{2} \sum_{j=1}^{2} \left(\delta_{j,1} - \delta_{j,-1} \right) \hat{a}^{\dagger}_{j} \hat{a}_{j},
\end{eqnarray}
to recover the cross-cavity QRM in Eq.(\ref{eq:CCQRM}) with parameters,
\begin{eqnarray}
\omega_{0} &=& - \frac{1}{2} \left( \delta_{1,-1} + \delta_{1,1} \right) = - \frac{1}{2} \left( \delta_{2,-1} + \delta_{2,1} \right) , \\
\omega_{j} &=& \frac{1}{2} \left( \delta_{j,-1} - \delta_{j,1} \right), \\
g_{j} &=& \eta_{j,\pm 1} \Omega_{j,\pm 1}  e^{- \frac{1}{2}\vert \eta_{j,\pm 1} \vert^{2}}.
\end{eqnarray}
Note, the definition of the qubit energy gap $\omega_{0}$ imposes the restriction that the sum of sideband driving field detunings for each mode must be equal, $ \delta_{1,-1} + \delta_{1,1}  = \delta_{2,-1} + \delta_{2,1}$.
Nevertheless, this gives us enough freedom to realize the cross-cavity QRM in a vast range of parameter sets.

\section{Spectra and eigenstates}

The feasibility of a quantum simulation of the cross-cavity QRM gives us a reason to explore the solution of this model.
In particular, finding the eigenvalues and eigenstates for the closed system may simplify the study of dissipation in the two-level system, which is a necessity in order to compare with experimental measurements.

We will first diagonalize our cross-cavity QRM in the two-level system basis following a Fulton-Gouterman approach.
For this, we rewrite the cross-cavity QRM Hamiltonian in terms of the qubit raising and lowering operators,
\begin{eqnarray}
\hat{H} = \frac{1}{2} \omega_{0} \hat{\sigma}_{0} + \sum_{j=1}^{2} \omega_{j} \hat{a}_{j}^{\dagger} \hat{a}_{j} + \left[  g_{1} \left( \hat{a}_{1}^{\dagger} + \hat{a}_{1} \right) - i g_{2} \left( \hat{a}_{2}^{\dagger} + \hat{a}_{2} \right) \right] \hat{\sigma}_{+} + h.c.,
\end{eqnarray}
and, in order to avoid imaginary couplings, we will perform a $\pi/2$ rotation around the second field photon number, $\hat{a}_{2}^{\dagger} \hat{a}_{2}$,
\begin{eqnarray}
\hat{H}_{R} = \frac{1}{2} \omega_{0} \hat{\sigma}_{0} + \sum_{j=1}^{2} \omega_{j} \hat{a}_{j}^{\dagger} \hat{a}_{j} + \left[  g_{1} \left( \hat{a}_{1}^{\dagger} + \hat{a}_{1} \right) + g_{2} \left( \hat{a}_{2}^{\dagger} - \hat{a}_{2} \right) \right] \hat{\sigma}_{+} + h.c.
\end{eqnarray}
Now, we can use Schwinger two-boson representation of $SU(2)$ and effect a $\pi/2$ rotation around $\hat{J}_{y} = -i (\hat{a}_{1}^{\dagger}\hat{a}_{2}-\hat{a}_{1}\hat{a}_{2}^{\dagger})/2$, $
\hat{D}_{\hat{J}_{y}}(\theta)=e^{ \frac{\theta}{2} \left(\hat{a}_{1}^{\dagger}\hat{a}_{2}-\hat{a}_{1}\hat{a}_{2}^{\dagger}\right)}
$ with $\tan \theta/2 = g_2/g_1$, to obtain the following,
\begin{eqnarray}
\hat{H}_{D} &=& \frac{1}{2} \omega_{0} \hat{\sigma}_{0} + \sum_{j=1}^{2} \Omega_{j} \hat{a}_{j}^{\dagger} \hat{a}_{j} + \gamma \left( \hat{a}_{1}^{\dagger} \hat{a}_{2} + \hat{a}_{1} \hat{a}_{2}^{\dagger}  \right) + \nonumber \\
&& + \frac{1}{\sqrt{g_{1}^{2} + g_{2}^{2} }}\left[  g_{1}^2 \left( \hat{a}_{1}^{\dagger} + \hat{a}_{1} \right) - g_{1} g_{2} \left( \hat{a}_{2}^{\dagger} + \hat{a}_{2} \right) \right] \hat{\sigma}_{1} + \nonumber \\
&&  + \frac{i}{\sqrt{g_{1}^{2} + g_{2}^{2} }}\left[  g_{2}^2 \left( \hat{a}_{1}^{\dagger} - \hat{a}_{1} \right) - g_{1} g_{2} \left( \hat{a}_{2}^{\dagger} - \hat{a}_{2} \right) \right] \hat{\sigma}_{2},
\end{eqnarray}
where we have used $\hat{\sigma}_{\pm} = (\hat{\sigma}_{1} \pm i \hat{\sigma}_{2} ) /2 $ and we have defined the following mixed field frequencies and beam-splitter parameter,
\begin{eqnarray}
\Omega_{1} = \frac{\omega_{1} g_{1}^2 + \omega_{2} g_{2}^2 }{g_{1}^{2} + g_{2}^{2}}, \quad \Omega_{2} = \frac{\omega_{2} g_{1}^2 + \omega_{1} g_{2}^2 }{g_{1}^{2} + g_{2}^{2}}, \quad \gamma = \frac{g_{1}g_{2}}{g_{1}^{2} + g_{2}^{2}}\left(\omega_{2} - \omega_{1} \right).
\end{eqnarray}
A rotation of $\pi/4$ around $\hat{\sigma}_{2}$ yields a Hamiltonian of the Fulton-Gouterman type \cite{Fulton1961p1059,Moroz2016},
\begin{equation}
\hat{H}_{FG} = \hat{A} \hat{\mathbbm{1}}  + \hat{B} \hat{\sigma}_{1} + \hat{C} \hat{\sigma}_{2} + \hat{D} \hat{\sigma}_{0},
\end{equation}
with auxiliary operators for our rotated cross-cavity QRM,
\begin{eqnarray}
\begin{array}{ll}
\hat{A}_{D} = \Omega_{1} \hat{a}^{\dagger}_{1} \hat{a}_{1} + \Omega_2 \hat{a}^{\dagger}_{2} \hat{a}_{2} + \gamma \left( \hat{a}_{1}^{\dagger} \hat{a}_{2} + \hat{a}_{1} \hat{a}_{2}^{\dagger}  \right) , \\
\hat{B}_{D} = -\frac{1}{2} \omega_{0}, \\
\hat{C}_{D} = \frac{i}{ \sqrt{ g_{1}^{2} + g_{2}^{2}} } \left[ g_{2}^{2} \left( \hat{a}_{1}^{\dagger} - \hat{a}_{1} \right) - g_{1} g_{2} \left( \hat{a}_{2}^{\dagger} - \hat{a}_{2} \right) \right] , \\
\hat{D}_{D} =  \frac{1}{ \sqrt{ g_{1}^{2} + g_{2}^{2}} } \left[ g_{1}^{2} \left( \hat{a}_{1}^{\dagger} + \hat{a}_{1} \right) - g_{1} g_{2} \left( \hat{a}_{2}^{\dagger} + \hat{a}_{2} \right) \right].
\end{array}
\end{eqnarray}
In order to diagonalize this in the two-level basis, we need an operator $\hat{R}$ such that,
\begin{eqnarray}
\left[ \hat{R}, \hat{A} \right] = \left[\hat{R}, \hat{B} \right]=\left\{ \hat{R}, \hat{C} \right\} = \left\{ \hat{R}, \hat{D} \right\} = 0.
\end{eqnarray}
We can choose the boson field parity,
\begin{eqnarray}
\hat{R} = \hat{\Pi}_{12} = e^{i \pi \sum_{j} \hat{a}_{j}^{\dagger} \hat{a}_{j}}.
\end{eqnarray}
to write a FG unitary transformation,
\begin{eqnarray}
\hat{U}_{FG} = \frac{1}{2 \sqrt{2}} \left[ \left( 1 + \hat{R} \right) \left( \hat{\sigma}_{0} + \hat{\sigma}_{1} \right) + \left( 1 - \hat{R} \right) \left( \hat{\mathbbm{1}} - i  \hat{\sigma}_{2} \right) \right],
\end{eqnarray}
that diagonalizes our cross-cavity QRM Hamiltonian in the qubit basis,
\begin{eqnarray}
\hat{H}_{FG}^{(D)} &=& \hat{U}_{FG} \hat{H}_{FG} \hat{U}_{FG}^{\dagger}, \\
&=& \left( \hat{A} + \hat{D} \right) \hat{\mathbbm{1}} + \left( \hat{B} - i \hat{C} \right) \hat{R} \hat{\sigma}_{0}.
\end{eqnarray}
Here, the diagonal form of our Hamiltonian in the qubit basis is
\begin{eqnarray}
\hat{H}_{D}^{(D)} &=& \hat{H}_{D}^{(+)} \vert e \rangle\langle e \vert + \hat{H}_{D}^{(-)} \vert g \rangle\langle g \vert,
\end{eqnarray}
with the auxiliary Hamiltonians in terms of just the two field modes,
\begin{eqnarray}
\hat{H}_{D}^{(\pm)} &=& \Omega_{1} \hat{a}_{1}^{\dagger} \hat{a}_{1} + \Omega_{2} \hat{a}_{2}^{\dagger} \hat{a}_{2} + \gamma \left( \hat{a}_{1}^{\dagger} \hat{a}_{2} + \hat{a}_{1} \hat{a}_{2}^{\dagger}  \right) \mp \frac{1}{2} \omega_{0} \hat{\Pi}_{12} + \nonumber \\
&& + \frac{1}{\sqrt{g_{1}^2 + g_{2}^{2}}}  \left[ \hat{a}_{1}^{\dagger} \left( g_{1}^{2} \pm g_{2}^2 \hat{\Pi}_{12} \right) + \left( g_{1}^{2} \pm g_{2}^2 \hat{\Pi}_{12} \right) \hat{a}_{1}   \right] + \nonumber \\
&& - \frac{g_{1} g_{2} }{\sqrt{g_{1}^2 + g_{2}^{2}}}  \left[  \hat{a}_{2}^{\dagger} \left( 1 \pm \hat{\Pi}_{12} \right)  +  \left( 1 \pm  \hat{\Pi}_{12} \right)  \hat{a}_{2}  \right].
\end{eqnarray}
These field Hamiltonians describe two driven boson fields interacting through a beam-splitter coupling.
The driving function depends on the parity of both fields.
These coupled and driven oscillators may be feasible of diagonalization using Bargmann representation \cite{Braak2011p100401} and we will address this in a future manuscript.
Here we are concerned with just a survey of the possibilities provided by our cross-cavity QRM.

\subsection{Fields with same frequencies and couplings}

We can provide a small amount of intuition if we consider two boson fields with the same frequency, $\omega_{1}=\omega_{2}=\omega$, and identical coupling strengths, $g_{1}=g_{2}=g$.
In this exceptional configuration, we can follow the aforementioned procedure step by step, with the slight deviation of introducing a rotation of $\omega_{0}/2$ around the operator,
\begin{eqnarray}
\hat{\eta} = - \hat{a}_{1}^{\dagger} \hat{a}_{1} + \hat{a}_{2}^{\dagger} \hat{a}_{2} + \frac{1}{2} \left( \hat{\sigma}_{0} + 1\right),
\end{eqnarray}
composed from the conserved operators from JC and anti-JC dynamics \cite{RodriguezLara2005p023811}.
This slight change allows us to recover an effective Hamiltonian where the first field mode is coupled to the qubit under anti-JC dynamics and the second under JC dynamics,
\begin{eqnarray}
\hat{H}_{exc}=\sum_{j} \delta_{j}\hat{a}_{j}^{\dagger}\hat{a}_{j}+\sqrt{2}g \left[(\hat{a}_{1}^{\dagger}-\hat{a}_{2})\hat{\sigma}_{+} + (\hat{a}_{1}-\hat{a}_{2}^{\dagger})\hat{\sigma}_{-}\right],
\end{eqnarray}
with field detunnings provided by the following expressions,
\begin{eqnarray}
\delta_{1}=\omega + \omega_{0}, \quad \delta_{2} = \omega - \omega_{0}.
\end{eqnarray}
This Hamiltonian will conserve the operator $\hat{\eta}$, $\left[ \hat{\eta}, \hat{H}_{exc} \right]=0$, which relates to Schwingers two-boson $SU(2)$ representation as $\hat{J}_{z} = ( \hat{a}_{1}^{\dagger} \hat{a}_{1} - \hat{a}_{2}^{\dagger} \hat{a}_{2})/2$; in other words, it is the difference between the qubit and the two-boson $SU(2)$ population difference.
If we choose the mean value of operator $\hat{\eta}$ to partition the corresponding Hilbert space, we will finish with infinite dimensional subspaces for each $\langle \hat{\eta} \rangle = 0, \pm 1, \pm2, \ldots$
Note that either the original cross-cavity QRM simulation with degenerate field frequencies and balanced couplings or this effective aJC-JC Hamiltonian can be implemented in our trapped ion simulation. 
Following the rest of the procedure, we produce the auxiliary operators,
\begin{eqnarray}
\begin{array}{ll}
\hat{A}_{exc} = \delta_{1} \hat{a}^{\dagger}_{1} \hat{a}_{1} + \delta_2 \hat{a}^{\dagger}_{2} \hat{a}_{2}, &
\hat{B}_{exc} = 0, \\
\hat{C}_{exc} = i \frac{1}{\sqrt{2}} g (\hat{a}_{1}^{\dagger} - \hat{a}_{1} + \hat{a}_{2}^{\dagger} -\hat{a}_{2}), &
\hat{D}_{exc} = \frac{1}{\sqrt{2}} g (\hat{a}_{1}^{\dagger} + \hat{a}_{1} - \hat{a}_{2}^{\dagger} -\hat{a}_{2}).
\end{array}
\end{eqnarray}
Here, the diagonal form of our exceptional Hamiltonian in the qubit basis is
\begin{eqnarray}
\hat{H}_{exc}^{(D)} &=& \hat{H}_{exc}^{(+)} \vert e \rangle\langle e \vert + \hat{H}_{exc}^{(-)} \vert g \rangle\langle g \vert,
\end{eqnarray}
with the auxiliary field Hamiltonians,
\begin{eqnarray}
\hat{H}_{exc}^{(\pm)} &=& \sum_{j=1}^{2} \delta_{j} \hat{a}_{j}^{\dagger} \hat{a}_{j} +   g \left( \hat{A}_{j}^{\dagger} + \hat{A}_{j} \right),
\end{eqnarray}
where we have defined nonlinear parity deformed operators,
\begin{eqnarray}
\hat{A}_{j} = - (-1)^{j}  \frac{1}{\sqrt{2}} \hat{a}_{j} \left(1 \pm (-1)^{j} \hat{\Pi}_{12} \right) , \\
\hat{A}_{j}^{\dagger} = - (-1)^{j}  \frac{1}{\sqrt{2}} \left(1 \pm (-1)^{j} \hat{\Pi}_{12} \right) \hat{a}_{j}^{\dagger} ,
\end{eqnarray}
that realize a Wigner-Heisenberg algebra \cite{Wigner1950p711,Yang1951p788},
\begin{eqnarray}
\left[ \hat{a}_{j}^{\dagger} \hat{a}_{j},\hat{A}_{j} \right] = - \hat{A}_{j},  \quad \left[ \hat{a}_{j}^{\dagger} \hat{a}_{j}, \hat{A}_{j}^{\dagger} \right] =  \hat{A}_{j}^{\dagger}, \nonumber \\
\left[ \hat{A}_{j}, \hat{A}_{j}^{\dagger} \right] = 1 \mp (-1)^{j} \left(2 \hat{a}_{j}^{\dagger} \hat{a}_{j} + 1 \right) \hat{\Pi}_{12},
\end{eqnarray}
for each boson field mode, note that these are not mutually compatible algebras, 
\begin{eqnarray}
\left[\hat{A}_{1}, \hat{A}_{2} \right] = \mp 2 \hat{\Pi}_{12} \hat{a}_{1} \hat{a}_{2}  , \qquad
\left[\hat{A}_{1}^{\dagger}, \hat{A}_{2} \right] = 0 .
\end{eqnarray}

\begin{figure}
	\centering \includegraphics[scale=1]{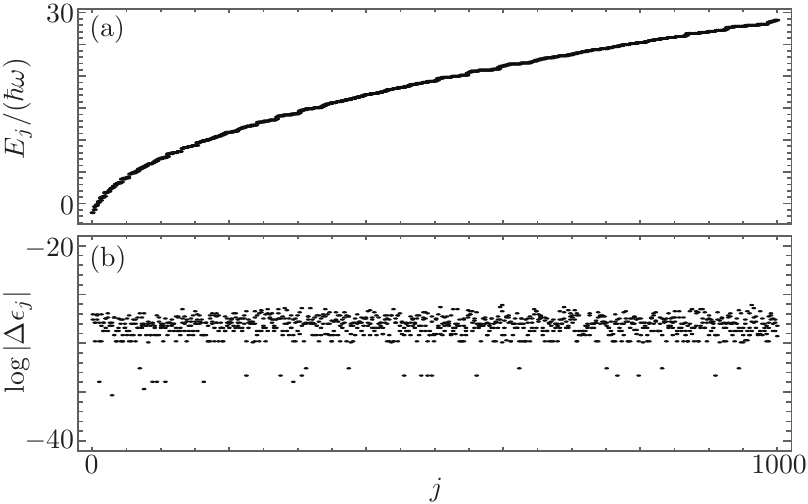}
	\caption{ (a) A thousand scaled energies of the spectra and (b) logarithm of the absolute value of the relative error between the numeric results, $\Delta \epsilon_{j} = E_{j}^{ccQRM} - E_{j}^{FG}$, for the cross-cavity QRM in the USC regime with $\omega_{1} = \omega_{2} = \omega_{0} = g_{1} = g_{2} = \omega$.} \label{fig:Fig2}
\end{figure}

At this point, we can use the two auxiliary nonlinear parity deformed oscillators to calculate the spectrum of our cross-cavity QRM for degenerate field frequencies and balanced couplings.
Figure \ref{fig:Fig2}(a) shows a thousand scaled energies, $E_{j}/\omega$, from the spectra of the cross-cavity QRM with equal field frequencies and balanced couplings in the ultra strong coupling (USC) regime, $\omega_{1}=\omega_{2}=g_{1}=g_{2}=\omega$.
For the sake of comparison, we also implemented a spectral solver for the cross-cavity QRM using up to a hundred photons in each of the boson fields.
Figure \ref{fig:Fig2}(b) shows the logarithm of the absolute value of the relative error between both numerical results, 
\begin{eqnarray}
\Delta \epsilon_{j} = E_{j}^{ccQRM} - E_{j}^{FG}.
\end{eqnarray}
One must be careful while rotating results to the same reference frame and ordering the combined spectra provided by the two auxiliary nonlinear parity deformed oscillators.
Note, our spectral solver using up to a hunderd photons on the fields delivers about five thousand converged eigensates that we can trust under a convergence criterion involving the information on the tail of the eigenstate that we will discuss later on.

\begin{figure}
\centering \includegraphics[scale=1]{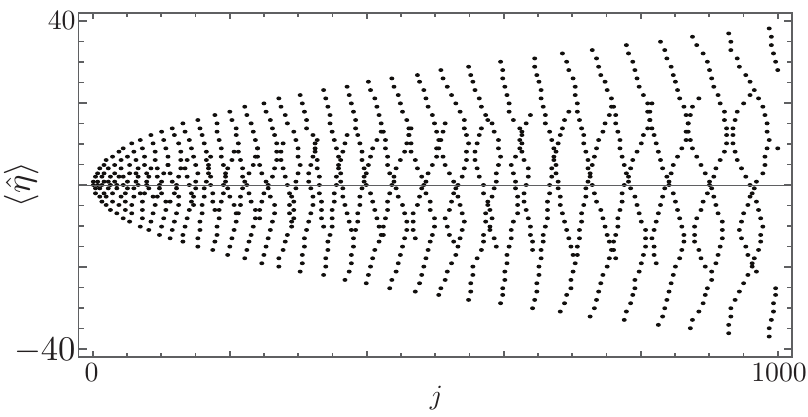}
\caption{ Mean value of the operator $\hat{\eta}$, $\langle \hat{\eta} \rangle$, for eigenstates in the frame of $\hat{H}_{exc}$ in the USC regime, $\omega_{1} = \omega_{2} = \omega_{0} = g_{1} = g_{2} = \omega$. Note the ordered lattice form due to the fact that $\hat{\eta}$ is conserved by $\hat{H}_{exc}$, $\left[ \hat{\eta}, \hat{H}_{exc} \right]=0$.} \label{fig:Fig3}
\end{figure}

As mentioned before, the effective JC-aJC Hamiltonian share eigenbasis with the operator $\hat{\eta}$.
Thus, the mean value of this operator for the numerical eigenstates of the nonlinear parity deformed oscillators, once transformed back into the frame of $\hat{H}_{exc}$, will form a so-called Peres lattice \cite{Peres1984p1711}, Fig. \ref{fig:Fig3}, where, for a given subspace defined by a constant value of $\langle \hat{\eta} \rangle =0 , \pm1, \pm2,\ldots$ in the frame of $\hat{H}_{exc}$, there will be an infinity number of eigenstates as shown in the figure.

\subsection{Weak coupling regime}

In the weak coupling regime (WCR), we can start from the cross-cavity QRM and move into the uncoupled rotating frame, $\hat{H}_{0}$, implement a RWA to neglect terms with frequency $\Delta_{j} = \omega_{j} + \omega_{0}$, and keep the terms with frequencies
\begin{eqnarray}
\delta_{j} &=& \omega_{j} - \omega_{0}.
\end{eqnarray}
Now, we move into a frame given provided by the free boson fields,
\begin{eqnarray}
\hat{H}_{n} &=& \sum_{j} \delta_{j} \hat{a}^{\dagger}_{j} \hat{a}_{j},
\end{eqnarray}
and, after implementing a rotation of $\pi/2$ around the frame defined by the number of bosons in the second mode $\hat{a}^{\dagger}_{2} \hat{a}_{2}$,  we obtain an effective Jaynes-Cummings model for each mode,
\begin{equation}
\hat{H}_{JC} = \delta_1 \hat{a}_{1}^{\dagger} \hat{a}_{1}  + \delta_2 \hat{a}_{2}^{\dagger} \hat{a}_{2} + g_{1}(\hat{a}_{1}^{\dagger}\hat{\sigma}_{-} + \hat{a}_{1}\hat{\sigma}_{+}) + g_{2}(\hat{a}_{2}^{\dagger}\hat{\sigma}_{-} + \hat{a}_{2}\hat{\sigma}_{+}).
\end{equation}
Again, either the original cross-cavity simulation with weak couplings or this effective JC model for the two fields can be implemented in our trapped ion simulation.
We can move this Hamiltonian into a FG form and end up with the auxiliary operators,
\begin{eqnarray}
\begin{array}{ll}
\hat{A}_{JC} = \delta_1 {\hat{n}}_1  + \delta_2 {\hat{n}}_2, &
\hat{B}_{JC} = 0, \\
\hat{C}_{JC} =  \sum_{j} \frac{i g_{j}}{2} \left(\hat{a}_{j} - \hat{a}_{j}^{\dagger} \right), &
\hat{D}_{JC} = \sum_{j} \frac{g_{j}}{2} \left(\hat{a}_{j} + \hat{a}_{j}^{\dagger} \right),
\end{array}
\end{eqnarray}
that lead to the boson field parity as the auxiliary operator, $\hat{R} = \hat{\Pi}_{12}$.
Finally, we obtain a diagonalized weak-coupling Hamiltonian,
\begin{eqnarray}
\hat{H}_{JC}^{(D)} = \hat{H}_{JC}^{(+)} \vert e \rangle\langle e \vert + \hat{H}_{JC}^{(-)} \vert g \rangle\langle g \vert,
\end{eqnarray}
where
\begin{eqnarray}
\hat{H}_{JC}^{(\pm)} = \sum_{j} \delta_{j} \hat{a}_{j}^{\dagger} \hat{a}_{j} + \frac{1}{2} g_{j} \left[ \hat{a}_{j}^{\dagger} \left( 1 \mp \hat{\Pi}_{12} \right) + \hat{a}_{j} \left( 1 \pm \hat{\Pi}_{12} \right) \right].
\end{eqnarray}

\begin{figure}
\centering \includegraphics[scale=1]{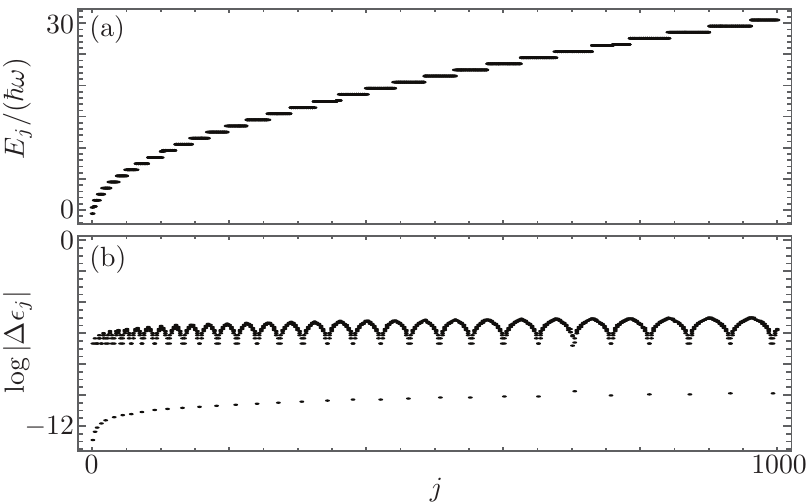}
\caption{ (a) A thousand scaled energies of the spectra and (b) logarithm of the absolute value of the relative error between the numeric results, $\Delta \epsilon_{j} = E_{j}^{ccQRM} - E_{j}^{FG}$, for the cross-cavity QRM in the weak coupling regime, $\omega_{1} = \omega_{2} = \omega_{0}= \omega$, $g_{1} = 0.001 \omega$ and $g_{2} = 0.002\omega$.} \label{fig:Fig4}
\end{figure}

\begin{figure}
\centering \includegraphics[scale=1]{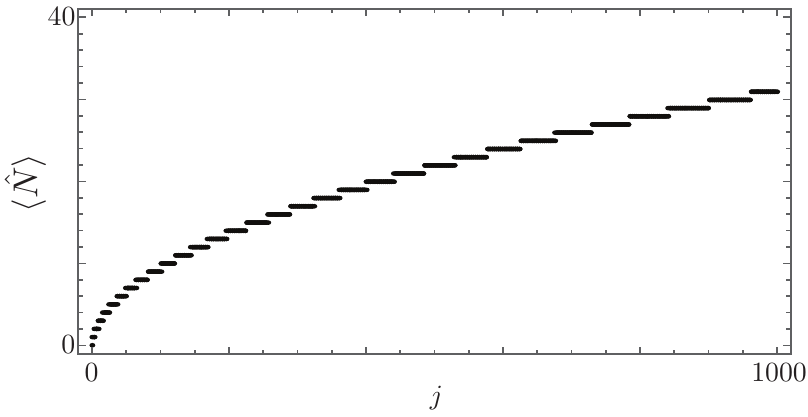}
\caption{ Mean value of the excitation number, $\langle \hat{N} \rangle$, for the eigenstates of the cross-cavity QRM in the weak coupling regime, $\omega_{1} = \omega_{2} = \omega_{0}= \omega$, $g_{1} = 0.001 \omega$ and $g_{2} = 0.002\omega$. Note the ordered lattice form due to the fact that $\hat{N}$ is conserved by $\hat{H}_{JC}$, $\left[ \hat{N}, \hat{H}_{JC} \right]=0$.} \label{fig:Fig5}
\end{figure}

Figure \ref{fig:Fig4}(a) shows a thousand energies of the spectra for the cross-cavity QRM with parameters in the weak coupling regime and Fig. \ref{fig:Fig4}(b) shows the logarith of the absolute value of the relative error between the numeric and analytic spectra.
Again, care must be exerted when ordering the analytic eigenvalues.
Furthermore, the effective JC model conserves the excitation number,
\begin{eqnarray}
\hat{N} = \hat{a}_{1}^{\dagger} \hat{a}_{1} + \hat{a}_{2}^{\dagger} \hat{a}_{2} + \frac{1}{2} \left( \hat{\sigma}_{0} + 1 \right),
\end{eqnarray}
such that $\left[ \hat{N}, \hat{H}_{JC} \right] = 0$. 
Thus, we can partition the corresponding Hilbert space into subspaces of dimension $2\langle \hat{N} \rangle+1$ for each an every average value of the excitation number $\langle \hat{N} \rangle = 0,1,2, \dots$ This can be observed in the structured Peres lattice for the mean value of the excitation number of the eigenstates, Fig. \ref{fig:Fig5}.

\subsection{Deep-strong coupling regime}

\begin{figure}
\centering \includegraphics[scale=1]{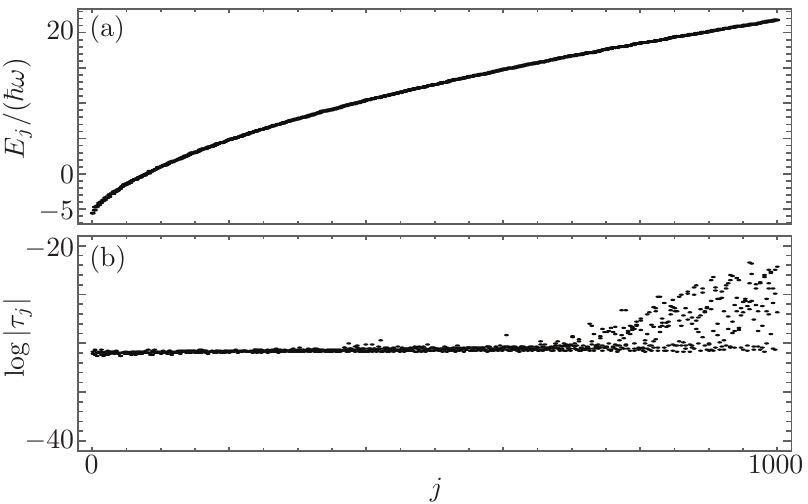}
\caption{ (a) A thousand scaled energies of the spectra and (b) logarithm of the absolute value of the squared root sum of the corresponding eigesntate tail for the cross-cavity QRM in the deep strong coupling regime, $\omega_{1} = \omega_{2} = \omega_{0}= \omega$, $g_{1} = 2 \omega$ and $g_{2} = 2.3 \omega$.} \label{fig:Fig6}
\end{figure}

\begin{figure}
\centering \includegraphics[scale=1]{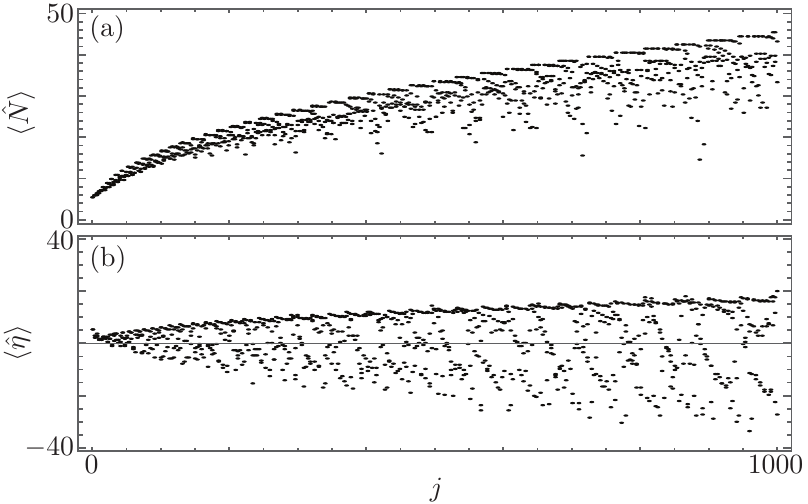}
\caption{ Mean value of the (a) excitation number, $\langle \hat{N} \rangle$, and (b) the operator, $\langle \hat{\eta} \rangle$, for the eigenstates of the cross-cavity QRM in the deep strong coupling regime, $\omega_{1} = \omega_{2} = \omega_{0}= \omega$, $g_{1} = 2 \omega$ and $g_{2} = 2.3 \omega$.} \label{fig:Fig7}
\end{figure}

In order to provide a solution for the cross-cavity QRM beyond the weak-coupling regime, we implement a brute force solver in a subspace allowing up to a hundred bosons in each field mode.
We will take as convergence measure the information content of the eigenstate tail \cite{BastarracheaMagnani2014p032101},
\begin{eqnarray}
\vert \tau_{j} \vert = \sqrt{ \sum_{k=k_{m}} \vert c_{k}^{(j)} \vert ^{2} },
\end{eqnarray}
where the complex number $c_{k}^{(j)}$ is the $k$-th element of the $j$-th eigenstate of the Hamiltonian expressed in the standard basis of the truncated subspace, either $\mathcal{H} = \mathcal{H}_{1} \otimes \mathcal{H}_{2} \otimes \mathcal{H}_{q}$ for the cross-cavity QRM or $\mathcal{H} = \mathcal{H}_{1} \otimes \mathcal{H}_{2}$ for the two nonlinear parity deformed driven oscillators, and we take the tail of the eigenstate as the last quarter of the truncated amplitudes, $ k_{m} = \lceil 3 \dim \mathcal{H}  / 4\rceil$ where the operation $\lceil x \rceil$ rounds up $x$ to the next integer.
Figure \ref{fig:Fig6} shows a thousand scaled energies and the logarithm of the information content of the eigenstate tail for the corresponding eigenstate for a cross-cavity QRM in the DSC regime with parameter values $\omega_{1}=\omega_{2}=\omega_{0}=\omega$, $g_{1}= 2 \omega$, and $g_{2} = 2.3 \omega$.
In the DSC regime, we lack any knowledge about the constants of motion, thus we will get disordered Peres lattices of the excitation number, Fig. \ref{fig:Fig7}(a), or the operator $\hat{\eta}$, Fig. \ref{fig:Fig7}(b), instead of the ordered lattices we obtained in the exceptional case of equal field frequencies and couplings or the weak-coupling regime, in that order.

\section{Conclusion}

We have proposed a cross-cavity quantum Rabi model, where a single two-level system interacts with two orthogonal boson fields under minimal coupling and the long-wavelength approximation, and shown it is feasible of experimental quantum simulation in the trapped ion quantum electrodynamics platform.

We diagonalized our model in the two-level basis following a Fulton-Gouterman approach and showed that the model reduces to a Hamiltonian where one boson field is coupled to the qubit following the Jaynes-Cummings model and the other boson field follows anti-Jaynes-Cummings coupling for all given coupling parameters in the particular case of fields with same frequencies and couplings.
In this peculiar regime, our model is equivalent to two parity deformed driven oscillators coupled via a beam splitter and conserves an operator that provides the total excitation number of the qubit and one field minus the excitation number of the other field.

For the sake of completeness, we calculated the spectra in the weak coupling regime and in the deep strong coupling regime but leave a detailed study including the influence of parity, which is conserved,  partial $SU(2)$ symmetry, and analytic solution via the now standard approach in Bargmann representation for future correspondence.

\section*{Acknowledgments}
C. Huerta Alderete acknowledges financial support through CONACYT master studies grant $\#331166$.

\section*{References}


\end{document}